\documentclass[a4paper,11pt]{article}
\usepackage{pos}
\usepackage{gensymb}
\usepackage{float}
\usepackage{amsthm}

\title{Compton-induced $\gamma$-ray cascade emissions in radio galaxy NGC 1275}
\ShortTitle{IC  cascade $\gamma$-rays in NGC 1275}

\newcommand{\unitU}{\ensuremath{\mathrm{erg}\cdot\mathrm{cm}^{-3}}}
\newcommand{\unitLum}{\ensuremath{\mathrm{erg}\cdot\mathrm{s}^{-1}}}

\newcommand{\unitRBLR}{\ensuremath{\mathrm{R}_{BLR}}}
\newcommand{\unitMsol}{\ensuremath{\mathrm{M}_\odot}}

\author*[a]{Mfuphi Ntshatsha}
\author[b]{Markus B{\"o}ttcher}
\author[a,c,d]{Soebur Razzaque}

\affiliation[a]{Centre for Astro-Particle Physics (CAPP) and Department of Physics, University of Johannesburg,\\
Auckland Park 2006, Johannesburg, South Africa}

\affiliation[b]{Centre for Space Research, North-West University, \\
Potchefstroom, 2520, South Africa}
\affiliation[c]{Department of Physics, The George Washington University,\\
Washington, DC 20052, USA}
\affiliation[d] {National Institute for Theoretical and Computational Sciences (NITheCS),\\
Private Bag X1, Matieland, South Africa}

\emailAdd{mfuphin95@gmail.com}
\emailAdd{Markus.Bottcher@nwu.ac.za}
\emailAdd{srazzaque@uj.ac.za}

\abstract{
Among active galactic nuclei (AGNi), blazars are the brightest emitters of high-energy (HE, $E \geq 100$ \,MeV) to very-high-energy (VHE, $E \geq 100$ \,GeV) $\gamma$-rays from their jets. Radio galaxies, being the misaligned parent population of the blazar class, were historically not detected at these frequencies. However, advances in experiments and observatories have led to their detection in the HE–VHE $\gamma$-ray band. In this work, we leverage and refine a Monte-Carlo photon and electron-positron (e$^\pm$) pair tracking code in the AGN environment of the radio galaxy NGC 1275. In the code, we consider the isotropic broad-line region (BLR) and anisotropic Shakura-Sunyaev (SS) accretion disk radiation fields, with mild magnetic fields in the AGN environment. We find that cascade $\gamma$-rays from inverse-Compton scattering by relativistic e$^\pm$ pairs of these external radiation fields can explain the \emph{Fermi} Large Area Telescope’s (LAT) and Major Atmospheric Cherenkov Experiment's observations from the radio galaxy NGC 1275. We present a set of plausible parameters obtained from the code by fitting the source’s spectral energy distribution (SED) during flaring events reported during the period December 2022 to January 2023.
}

\FullConference{High Energy Astrophysics in Southern Africa (HEASA2025)\\
16-20 September, 2025\\
University of Johannesburg, South Africa\\}

\begin{document}
\maketitle

\section{Introduction}
\label{sec:intro}
Some galaxies accrete matter onto their supermassive black holes (SMBHs), forming accretion disks, sometimes accompanied by the expulsion of relativistic jets, emitting at all frequencies of the electromagnetic spectrum. When an AGN is oriented with its relativistic jet aligned with our line of sight, it is classified as a blazar. Of all AGNi, blazars are the brightest $\gamma$-ray emitters. This is due to their emissions being Doppler boosted as a consequence of the relativistic motion of the jet. This is supported by emission models, where high Doppler factors are required to explain their SEDs. According to the unified scheme of AGN \cite{urry1995unified}, radio galaxies are the misaligned parent population of the blazars. This would make them intrinsic producers of HE--VHE $\gamma$-rays. However, these $\gamma$-rays cannot be observed given that we do not have a head-on view of the jet in radio galaxies. Improved sensitivity in $\gamma$-ray observatories has led to the detection of \,GeV and even \,TeV $\gamma$-rays from some radio galaxies. This is challenging to explain given that we cannot rely on the Doppler enhancement of this emission. In these proceedings, we address this by leveraging the works of \cite{roustazadeh2010very, roustazadeh2011very, roustazadeh2012synchrotron}. In their works, the authors developed a Monte-Carlo (MC) code that propagates individual $\gamma$-ray photons in an AGN environment and follows the development of inverse-Compton (IC) $\gamma$-ray and e$^\pm$ cascades in three dimensions (3D). We refined this code by including a standard thin Shakura-Sunyaev accretion disk model \cite{shakura1973black}. In a photon-processing routine, photon outputs from this code have been compared with the broadband SED data points of radio galaxy NGC 1275 \cite{godambe2024very}. 

\section{Code setup}
\label{sec:scode_etup}
The MC code propagates individual $\gamma$-rays through an AGN environment, initially produced in the jet. The \emph{Fermi}-LAT spectra of many AGNi can be fit with a power-law \cite{abdo2009Fermi, abdo2010fermi, abdo2011fermi}. Therefore, without presuming to know the origin of the primary $\gamma$-rays, we assume they have a power-law spectrum. In the cascade code, a primary $\gamma$-ray's dimensionless energy $\epsilon_\gamma \equiv E_\gamma / (m_e c^2)$ is drawn from a power-law distribution $(\propto \epsilon_\gamma^{-\alpha})$ with power-law index $\alpha$, where the constants $m_e, c$ are the electron rest-mass and speed of light in vacuum, respectively. The photon initially travels in the forward jet direction and either escapes or is absorbed via $\gamma-\gamma$ pair production. Depending on the opacity, $\kappa_{\gamma\gamma}^{BLR}$ or $\kappa_{\gamma\gamma}^{d}$, the $\gamma$-ray can be either absorbed by a broad-line region (BLR) or accretion disk photon, respectively. The absorbing target photon is chosen by drawing a uniform random number $\zeta_{\gamma\gamma} \in (0,1)$ and comparing it to the fraction $\kappa_{\gamma\gamma}^d / \kappa_{\gamma\gamma}^{tot}$, where $\kappa_{\gamma\gamma}^{tot} = \kappa_{\gamma\gamma}^{BLR} + \kappa_{\gamma\gamma}^d$. If $\zeta_{\gamma\gamma} < \kappa_{\gamma\gamma}^d / \kappa_{\gamma\gamma}^{tot}$, then the disk photon is the target photon otherwise it is the BLR photon, creating an e$^\pm$ pair. The pair initially travels in the direction of the absorbed $\gamma$-ray, and the particles are each deflected by the magnetic field. Along the deflected path, they emit synchrotron photons while IC scattering optical/ultraviolet (UV) photons from either the BLR or accretion disk. Similarly to selecting an absorbing photon, the target photon to be IC boosted to high energy is selected by calculating the e$^\pm$'s inverse-Compton mean-free path length $(\lambda_C^{BLR})^{-1}$ or $(\lambda_C^{d})^{-1}$ and comparing it to a random number $\zeta_C \in (0,1)$. The secondary $\gamma$-ray photon is then tracked in the same way as just described. This continues until the $\gamma$-ray or e$^\pm$ escapes the AGN environment or their energies $\epsilon < \epsilon_c, \gamma_\pm < \gamma_c$ fall below a threshold energy of interest set to $\epsilon_c = \gamma_c = 100$.

\section{Simulated parameters}
\label{sec:parameters}
To simulate an AGN, its parameters have to be input into the code. These are: its accretion disk luminosity ($L_d$), BLR size radius ($R_{BLR}$), primary $\gamma$-ray spectral index ($\alpha$), black hole mass ($m_{BH}$), magnetic field orientation ($\theta_B$) with respect to the jet axis, magnetic field strength ($B$), BLR energy density ($u_{BLR}$), and primary $\gamma$-ray origin $(x^\gamma_0, y^\gamma_0, z^\gamma_0)$ of which $z^\gamma_0$ is the injection height. The radio galaxy NGC 1275 is a type 1.5 Seyfert galaxy viewed at an angle constrained by \cite{godambe2024very} to be no more than 17\,\degree. Ref. \cite{godambe2024very} fit the SED of this source using a one-zone synchrotron self-Compton (SSC) model and obtained low Doppler factors ranging from 2 to 3.5. All calculations in the MC cascade code for this work are done in the AGN frame, thus the Doppler factor was not required. In all presented SEDs, the viewing angle is in the bin $10\,\degree < \theta <17\,\degree$, corresponding to the cosine angular bin $0.94 < \mu\equiv\cos\theta < 0.98$. To be consistent with the literature, we fixed the first five parameters to the literature-quoted values \cite{roustazadeh2010very, godambe2024very}. The varied parameters are $B, u_{BLR}, z^\gamma_0$ and their effects are discussed in Section~\ref{sec:results}. The values of the fixed parameters are tabulated in Table~\ref{tab:parameters}.

\begin{table}[h]
    \centering
    \begin{tabular}{c|c} \hline
         Parameter  &   Value                           \\\hline
         $L_d$      &   $1.88\times 10^{43}$\,\unitLum  \\
         $R_{BLR}$  &   $10^{16}$\,cm                   \\
         $\alpha$   &   2.5                             \\
         $m_{BH}$   &   $10^8$\,\unitMsol               \\
         $\theta_B$ &   11\,\degree                     \\\hline
    \end{tabular}
    \caption{Values of fixed parameters used in the simulation of NGC 1275.}
    \label{tab:parameters}
\end{table}

\section{Results and discussion}
\label{sec:results}
\begin{figure}[h]
    \centering
    \includegraphics[width=0.7\linewidth]{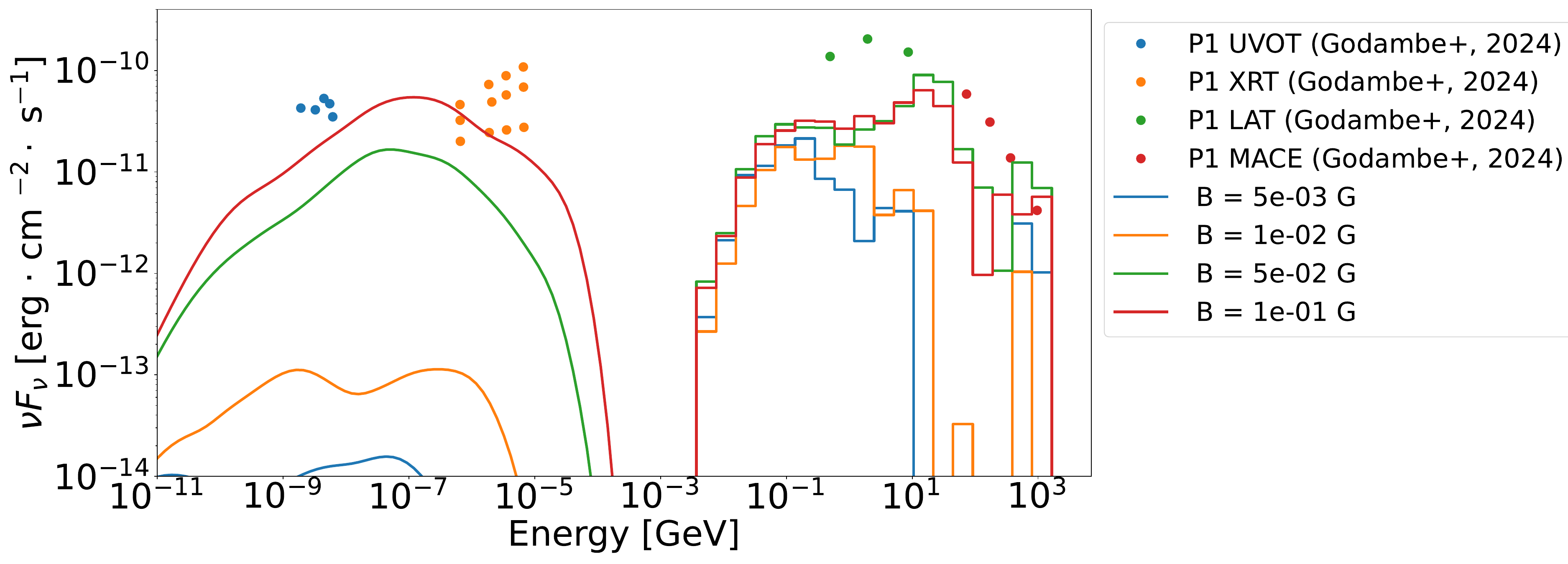}
    \caption{Cascade SED for different magnetic fields at BLR photon energy density of $50\times 10^{-3}$\,\unitU\ and injection height of 0.6\,\unitRBLR.}
    \label{fig:Bu50z00.6_varB}
\end{figure}

We discuss results from cascade SEDs, demonstrating the effects of varying three parameters, namely magnetic field strength, BLR energy density and primary $\gamma$-ray injection height. In these proceedings, we present SEDs constructed from the output of an MC cascade code which considers a homogeneous and isotropic BLR radiation field and an SS accretion disk radiation field. These SEDs were plotted against data published by \cite{godambe2024very}. In the period of December 2022 -- January 2023, the Major Atmospheric Cherenkov Experiment (MACE) detected flares from NGC 1275. Ref. \cite{godambe2024very} reported these findings and performed an analysis using simultaneous data from the \emph{Swift}-Ultraviolet and Optical Telescope (UVOT), \emph{Swift}-X-ray Telescope (XRT) and \emph{Fermi}-LAT to construct a broadband SED of NGC 1275. The \emph{blue, orange, green} and \emph{red} dots in the SED figures are \emph{Swift}-UVOT, \emph{Swift}-XRT, \emph{Fermi}-LAT and MACE data, respectively. The P1 data label in the legend of Figure~\ref{fig:Bu50z00.6_varB} is used as in \cite{godambe2024very}, representing the flare of the source during the night of 21 December 2022. X-ray data correspond to XRT observations on Modified Julian Dates (MJDs) 59933.9 and 59935.9 and the average flux of such days, as described in Figure 4 of \cite{godambe2024very}. We note that the source was not spatially resolved. The optical/UV and X-ray flux from this source is subject to contamination from the host galaxy, accretion-disk -- corona system, and possibly also the intracluster medium \cite{godambe2024very}. Therefore the optical/UV -- X-ray flux data are effectively upper limits for the cascade emission in these frequency bands.

To study the behaviour of cascade SED's response to magnetic field strength, we fixed $u_{BLR}$ and $z^\gamma_0$ in our simulations. In Figure~\ref{fig:Bu50z00.6_varB}, we show simulation runs with fixed $u_{BLR} = 50\times 10^{-3}$\,\unitU\ and $z^\gamma_0 = 0.6$\,\unitRBLR. As expected, the synchrotron peak flux rises according to the known scaling $\nu_{s,peak}F(\nu_{s,peak}) \propto B^2$. The behaviour of the IC flux is characterized by a rise with magnetic field strength, followed by a saturation at its maximum from few tens of milli-Gausses. This trend was also noted by \cite{roustazadeh2010very} and interpreted as the isotropization of the pairs by the transverse component of the magnetic field. Here, we also note that increasing the magnetic field has the effect of broadening the emission band of the IC component. The narrow $\gamma$-ray spectrum seen with low magnetic fields may be due to the increased gyration radius ($\propto \gamma_\pm B^{-1}$) inefficiently deflecting the pairs of energy $E_\pm = \gamma_\pm m_e c^2$. In this scenario, the vast majority of the highest energy cascade $\gamma$-rays are likely emitted in the forward direction, thus weak magnetic fields create the effect of truncating IC emission.

\begin{figure}[h]
    \centering
    \includegraphics[width=0.49\linewidth]{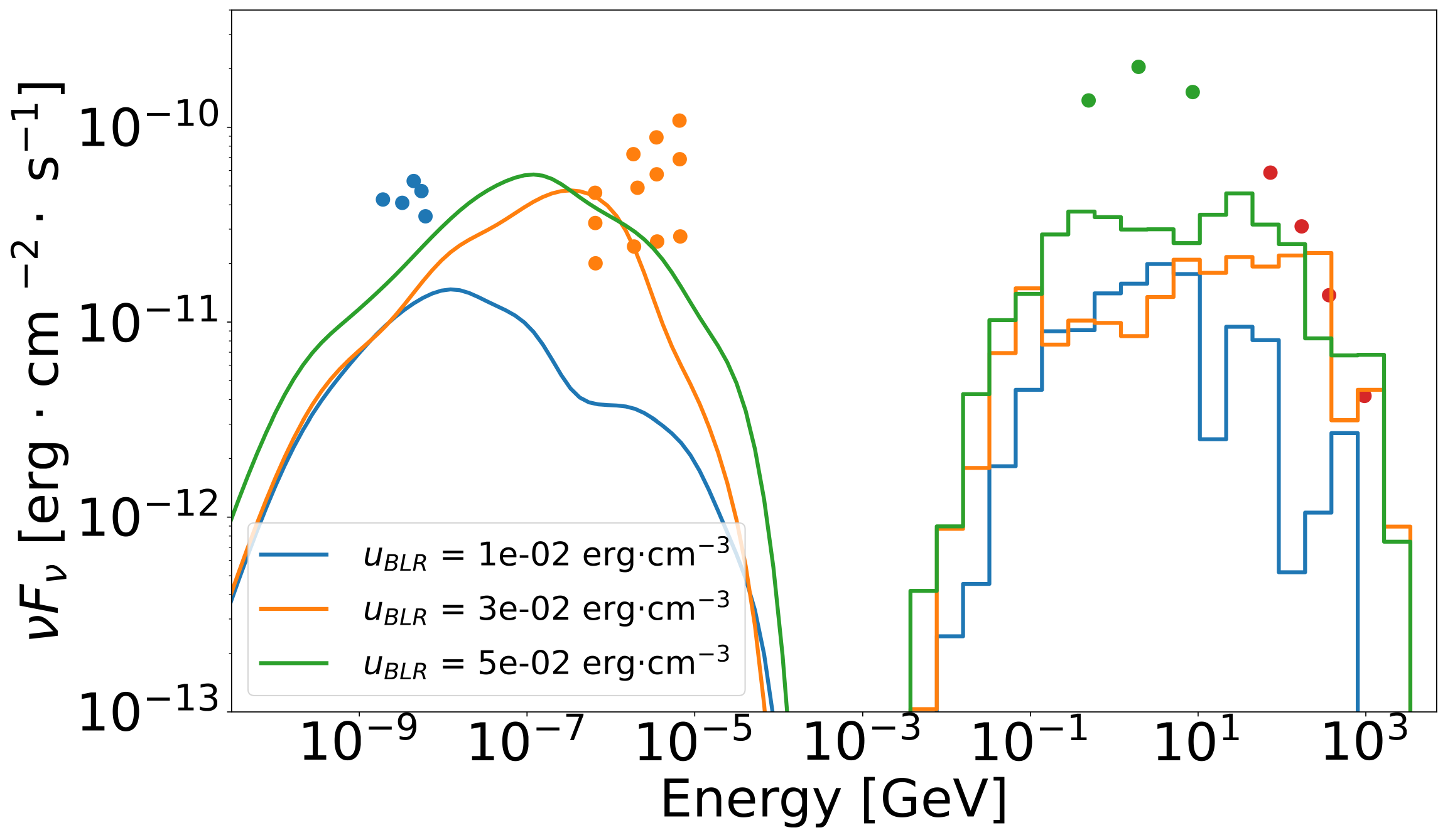}
    \includegraphics[width=0.49\linewidth]{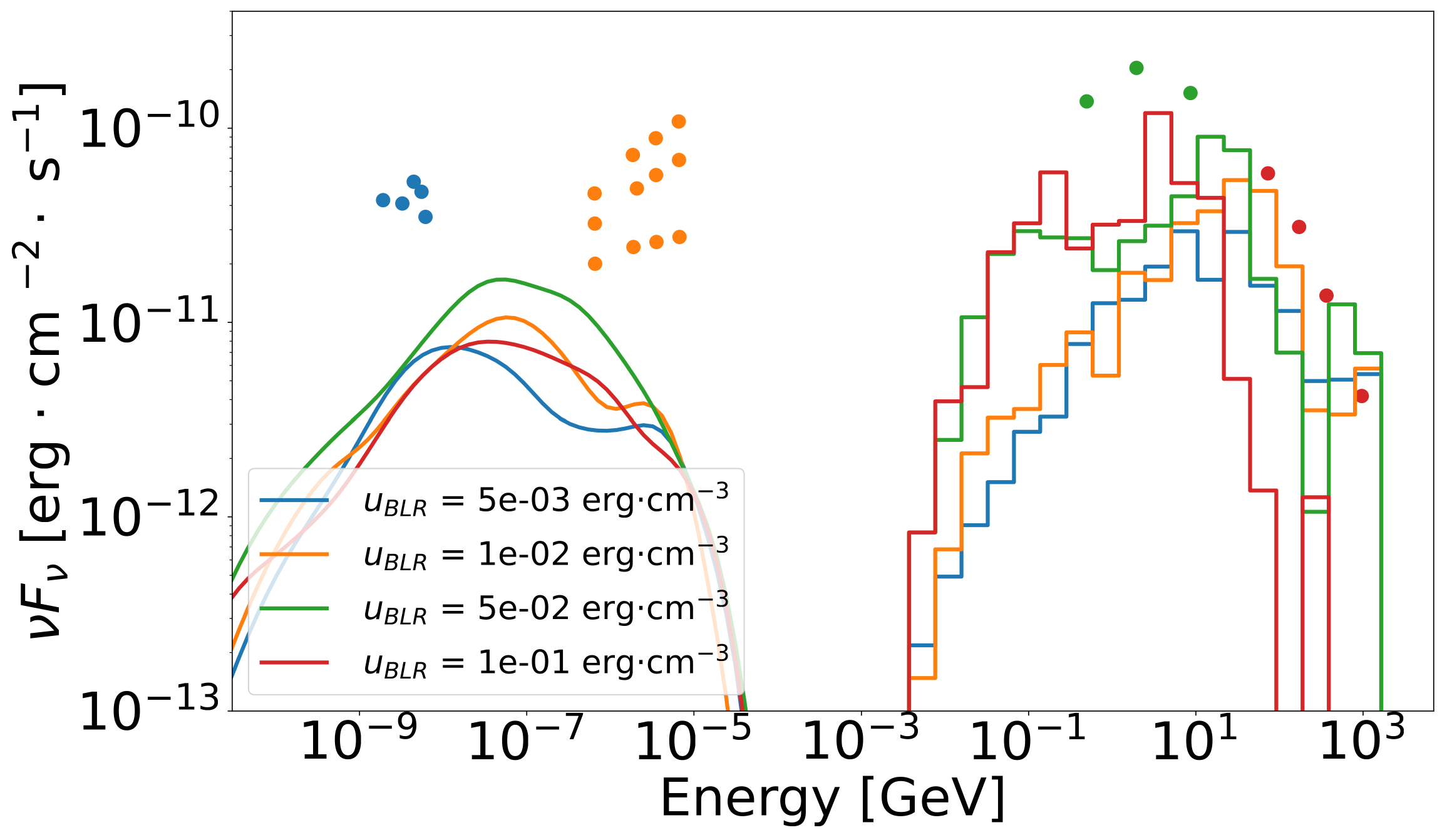}
    \caption{Cascade SED for different BLR energy densities. In the left panel the magnetic field is 100\,mG and the injection height is 0.8\,\unitRBLR, and 50\,mG and 0.6\,\unitRBLR\ in the right panel.}
    \label{fig:B100uz00.8_varu}
\end{figure}

To study the cascade SED across the full $\gamma$-ray range, we then fixed the magnetic field to 100\,mG and injected primary $\gamma$-rays at 0.8\,\unitRBLR, varying the $u_{BLR}$. These SEDs are shown in Figure~\ref{fig:B100uz00.8_varu}. Cascade emission increases with increasing $u_{BLR}$. A radiation field with energy density $u_{rad}$ permeating a region of radius $R_{rad}$ has a luminosity $L_{rad} = 4\pi c R_{rad}^2 u_{rad}$. Comparing the synchrotron luminosity $L_{syn}$ to the BLR luminosity $L_{BLR}$ in the same region size thus goes as the ratio of their energy densities $L_{syn}/L_{BLR} = u_{B}/u_{BLR}$. Initially, synchrotron emission increases with rising $u_{BLR}$. This is likely associated with an increase in the development of e$^\pm$ cascades. There is, however, a critical point where it becomes suppressed due to this ratio decreasing. This transition is shown in the left panel of Figure~\ref{fig:B100uz00.8_varu}, where simulation runs corresponding to BLR energy densities of $u_{BLR} = 50\times 10^{-3}$ and $30\times 10^{-3}$\,\unitU\ have comparable synchrotron peak fluxes. We note that from more simulation runs, further increases to $u_{BLR}$ led to the synchrotron emission dropping. For example in the right panel of Figure~\ref{fig:B100uz00.8_varu} at an intermediate injection height of 0.6\,\unitRBLR\ and a magnetic field of 50\,mG, we also simulated a $u_{BLR} = 100\times 10^{-3}$\,\unitU, corresponding to a BLR reprocessing 20\% of the disk luminosity. Here, the synchrotron emission has visibly dropped below that corresponding to $u_{BLR} = 50\times 10^{-3}$\,\unitU. This suggests that there is a limit to the BLR photon energy density beyond which the synchrotron emission is too far suppressed to be able to reproduce the observed optical/UV -- X-ray flux. Increasing $u_{BLR}$ raises the number of IC collissions, leading to a greater development of cascades. The figures demonstrate this with the rise in the IC flux. However, increasing $u_{BLR}$ also increases the BLR's opacity. In the right panel of Figure~\ref{fig:B100uz00.8_varu}, at $u_{BLR} = 100\times 10^{-3}$\,\unitU, \,TeV photons are severely attenuated.

\begin{figure}[t]
    \centering
    \includegraphics[width=0.49\linewidth]{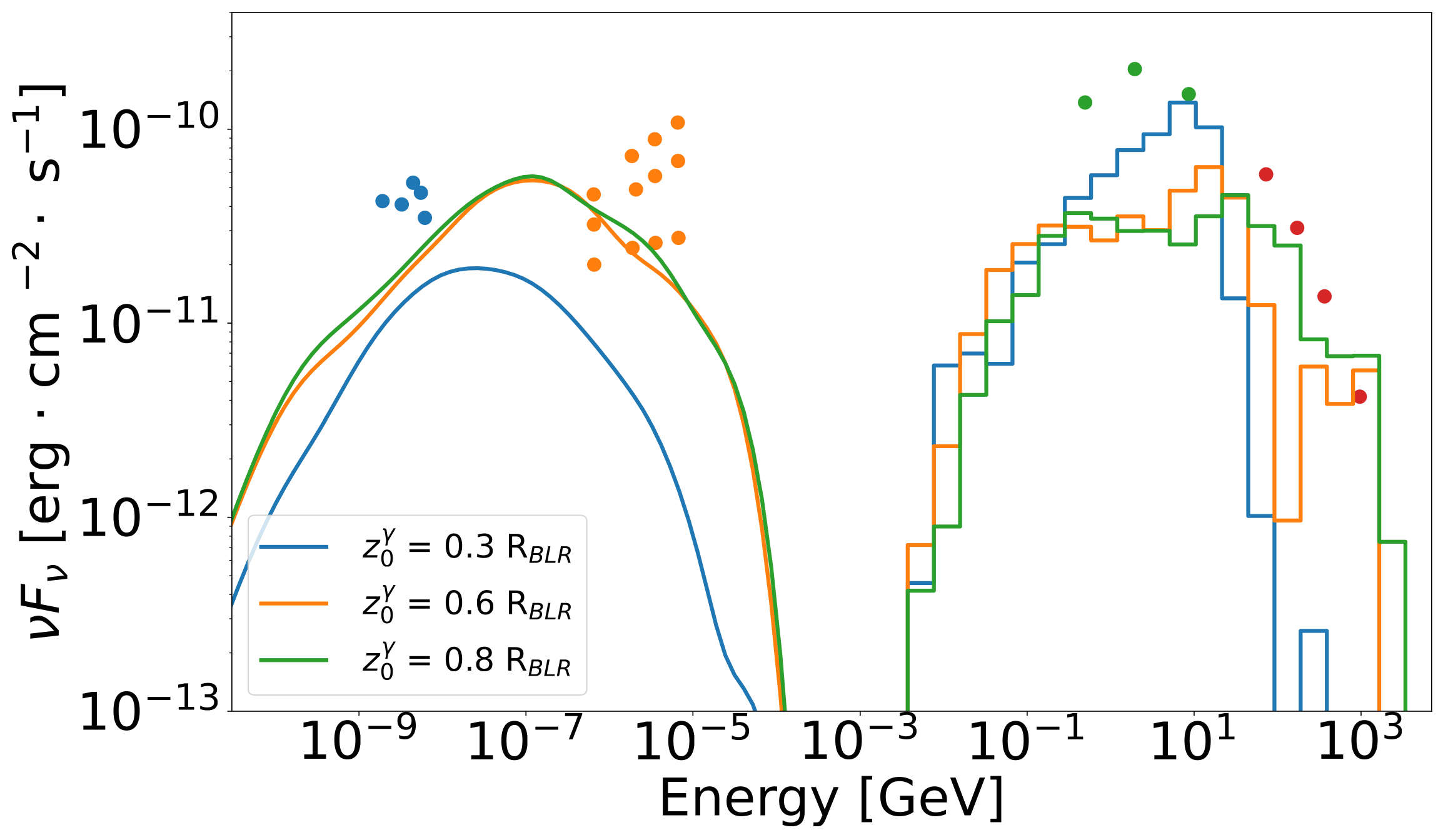}
    \includegraphics[width=0.49\linewidth]{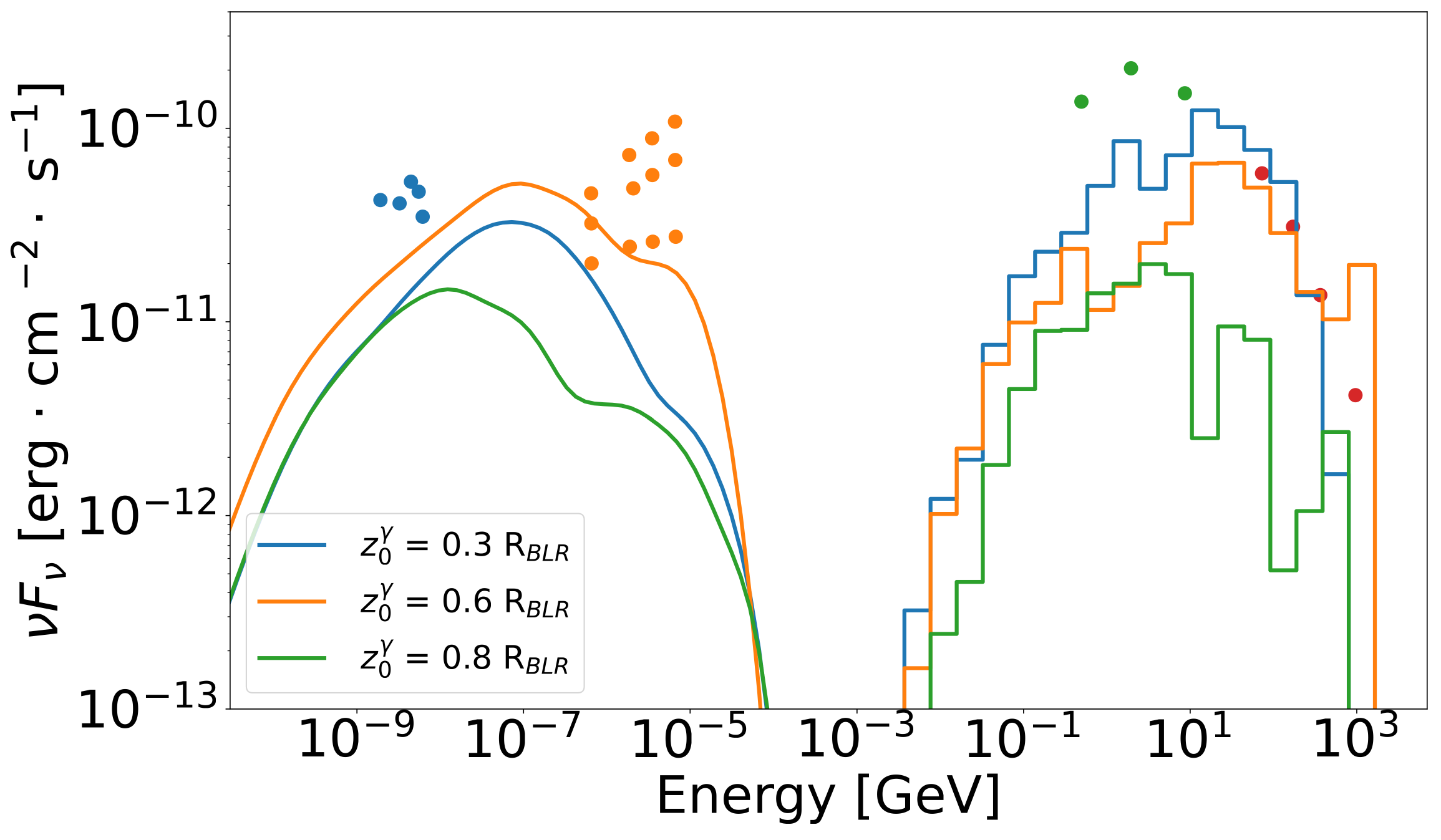}
    \caption{Cascade SED for different primary photon injection heights, where the magnetic field is 100\,mG and the BLR energy density is $u_{BLR} = 50\times 10^{-3}$\,\unitU\ in the left panel and $u_{BLR} = 10\times 10^{-3}$\,\unitU\ in the right panel.}
    \label{fig:B100u50z0_varR0}
\end{figure}

In addition to simulations with varied $B$ and $u_{BLR}$, we also changed the origin $z^\gamma_0$ of the injected $\gamma$-ray photons. Decreasing the injection height of primary photons has an effect similar to increasing $u_{BLR}$. $\gamma$-rays injected at heights closer to the BLR's outer edge are able to escape before suffering many, if any, IC collisions with the soft radiation fields.  Therefore, \,TeV $\gamma$-rays can more easily be observed as compared to injecting close to the base of the jet. By comparison, at the height of 0.3\,\unitRBLR, where at $u_{BLR} = 50\times 10^{-3}$\,\unitU, \,TeV emission is completely attenuated. This is similar to the right panel of Figure~\ref{fig:B100uz00.8_varu} where at $u_{BLR} = 100\times 10^{-3}$\,\unitU\ and intermediate height 0.6\,\unitRBLR, \,TeV $\gamma$-rays are also strongly absorbed. In the left panel of Figure~\ref{fig:B100u50z0_varR0}, we show simulation results with $u_{BLR} = 50\times 10^{-3}$\,\unitU. The synchrotron component of the SED for $z^\gamma_0 = 0.3$\,\unitRBLR\ is suppressed compared to injecting further downstream the jet. Also, \,TeV photons at this height are completely absorbed. In the right panel of Figure~\ref{fig:B100u50z0_varR0}, the BLR energy density is reduced to $10\times 10^{-3}$\,\unitU. With this $u_{BLR}$, \,TeV photons can escape even at the deepest simulated height ($z^\gamma_0 = 0.3$\,\unitRBLR).

\section{Conclusion}
\label{sec:conclusion}
By following the propagation of $\gamma$-rays in 3D, inverse-Compton supported $\gamma$-ray cascades can reliably explain the broadband emissions from AGNi viewed at large angles. The MC cascade code demonstrates this by especially not relying on the need for Doppler boosting of emissions in the AGN jets. We have thus far studied the effects of varying $B, u_{BLR}$ and $z^\gamma_0$ on the cascade SED. There seems to be some degeneracy between $u_{BLR}$ and $z^\gamma_0$, however. This highlights the interplay between the BLR and accretion disk photon fields, since the accretion disk photon field dominates at low $z^\gamma_0$. A magnetic field with a component perpendicular to the initial direction of primary $\gamma$-rays is a vital requirement to isotropize e$^\pm$ pairs about the AGN environment. At moderate BLR energy densities, the injection height of the primary $\gamma$-ray beam may also play a role in explaining \emph{Fermi}-LAT and MACE flux. While $\gamma$-rays injected close to the base of the jet may develop cascades that potentially reproduce the shape and flux levels of the \emph{Fermi}-LAT flux, those injected at intermediate to large heights may explain MACE flux levels. Although studies on the other parameters have not been fully investigated, the three presented here seem to require the need for primary $\gamma$-rays to be produced in at least two different zones.

\section{Acknowledgements}
This work would not have been possible without the availability of high-performance computational resources from the University of Johannesburg and the National Integrated CyberInfrastructure System.

\bibliographystyle{JHEP}
\bibliography{refs_mfp}
%

\end{document}